\documentclass[preprint,12pt]{elsarticle}
\pdfoutput=1 
\journal{Nuclear Physics B}

\usepackage[T1]{fontenc} 

\RequirePackage{flushend}
\usepackage{graphicx}
\usepackage{dcolumn}
\usepackage{bm}
\usepackage{amsmath}
\usepackage{multirow}
\usepackage{enumerate}
\usepackage{amsfonts}
\usepackage{ifthen}
\usepackage{psfrag}
\usepackage{slashed}
\usepackage{hyperref}
\usepackage{gensymb}
\usepackage[utf8]{inputenc}
\usepackage{color}
\usepackage{subfigure}
\usepackage{url}
\usepackage{amssymb}

\newcommand{\be}{\begin{equation}}
\newcommand{\ee}{\end{equation}}
\newcommand{\bea}{\begin{eqnarray}}
\newcommand{\eea}{\end{eqnarray}}

\begin{document}

\begin{frontmatter}

\title{Lepton Number Violating Electron Recoils in a $U(1)_{B-L}$ Model with Non-Standard Interactions}

\author[1,2]{Yugen Lin}
\ead{linyugen@ihep.ac.cn}
\author[1]{Yu Gao}
\ead{gaoyu@ihep.ac.cn}
\author[2,3]{Tianjun Li}
\ead{tli@itp.ac.cn}

\address[1]{Key Laboratory of Particle Astrophysics, Institute of High Energy Physics, Chinese Academy of Sciences, Beijing, 100049, China}
\address[2]{School of Physical Sciences, University of Chinese Academy of Sciences, Beijing, 100049, China}
\address[3]{CAS Key Laboratory of Theoretical Physics, Institute of Theoretical Physics, Chinese Academy of Sciences, Beijing 100190, China}




\begin{abstract}
We propose an $SU(3)_C\times SU(2)_L\times U(1)_{Y}\times U(1)_{B-L}$ model, in which the neutrino masses and mixings can be generated via Type-I seesaw mechanism after $U(1)_{B-L}$ breaking.  A light mediator emerges and enables non-standard interaction that violates the lepton number. We show that the non-standard neutrino interaction emerges in this model, and it can lead to low energy recoil events with the solar neutrino flux. Analyses are performed with the keV range electron recoil events at recent direct detection experiments, including XENON1T, PANDAX and XENONnT. 
Recent direct detection observations lead to upper bound on the combined coupling strength to electron and neutrino to $\sqrt{y'_\nu y_e} < 0.5 \times 10^{-6}$.
\end{abstract}


\end{frontmatter}

\section{Introduction}


Low-energy electron recoil received growing interest due to recent advances in direct detection with lowering thresholds ~\cite{XENON:2015gkh,DARWIN:2016hyl,DarkSide-20k:2017zyg,LZ:2018qzl,Schumann:2015cpa,SENSEI:2020dpa,SuperCDMS:2015eex}. Recent keV range electron-recoil observations at XENON1T~\cite{Aprile:2020tmw}, PANDAX~\cite{Zhou:2020bvf} and XENONnT~\cite{XENON:2022mpc}, together with improved scrutiny on experimental radiative backgrounds, offer high-quality data for probing any new physics that leads to low-energy electron recoils, such as low-mass dark matter~\cite{Essig:2011nj, Bringmann:2018cvk}, axions~\cite{Corsico:2014mpa, Giannotti:2017hny, Diaz:2019kim} and new interactions with neutrinos ~\cite{Cerdeno:2016sfi,Dutta:2017nht,AristizabalSierra:2017joc,Gonzalez-Garcia:2018dep}, etc.
With new data in the keV range, recent new physics studies with electron recoils include tests of
axion-like dark matter~\cite{Takahashi:2020bpq},
solar axions~\cite{DiLuzio:2020jjp,Gao:2020wer},
non-standard neutrino interactions with light mediators~\cite{Amaral:2020tga,Boehm:2020ltd,Bally:2020yid,AristizabalSierra:2020edu,Khan:2020vaf,An:2020bxd,Zu:2020idx,Lindner:2020kko,Khan:2020csx}, hidden photon dark matter~\cite{Alonso-Alvarez:2020cdv,Nakayama:2020ikz,Bloch:2020uzh},
warm or fast moving dark matter~\cite{Kannike:2020agf} (also see~\cite{Campos:2016gjh}),
boosted~\cite{Fornal:2020npv,Chen:2020gcl,Cao:2020bwd,Jho:2020sku, Jho:2021rmn},
inelastic or multi-component dark matter~\cite{Harigaya:2020ckz,Su:2020zny,Lee:2020wmh,Bramante:2020zos,Baryakhtar:2020rwy,He:2020sat},
decaying dark matter~\cite{Du:2020ybt,Choi:2020udy,Buch:2020mrg},
Migdal effect~\cite{Dey:2020sai},
luminous or shining dark matter~\cite{Bell:2020bes,Paz:2020pbc},
inverse Primakoff effect~\cite{Dent:2020jhf},
hydrogen decay~\cite{McKeen:2020vpf},
dark fluxes from accreting black holes~\cite{Cai:2020kfq},
as well as re-examining detector backgrounds~\cite{Robinson:2020gfu}, collider searches~\cite{Primulando:2020rdk}, neutrino magnetic moment~\cite{Chala:2020pbn}, stellar cooling~\cite{DeRocco:2020xdt, Dev:2020jkh} limits, etc.

In this paper, we propose a $SU(3)_C\times SU(2)_L\times U(1)_{Y}\times U(1)_{B-L}$ model to
derive low-energy electron recoils via non-standard neutrino-electron interactions with a light mediator.
After the $U(1)_{B-L}$ breaking, the neutrino masses and mixings can be generated via the Type I seesaw mechanism. In particular, a light mediator exists in the model, as well as the non-standard interactions between the light mediator and leptons which violate the lepton number. Such light mediator would enhance the electron recoil rates at low energy region. We show that significant constraints on the model's effective electron-neutrino couplings can be derived from the keV range electron recoil data from XENON1T, PANDAX and XENONnT.

\section{Model Setup}
\label{sect:Model}

The particles in the $SU(3)_C\times SU(2)_L\times U(1)_{Y}\times U(1)_{B-L}$ model follow the
conventional notations that the Standard Model (SM) quark doublets $Q_i$, right-handed
up-type quarks $U_i$, right-handed down-type quarks $D_i$, lepton doublets $L_i$, right-handed charged leptons $E_i$,
and right-handed neutrinos  $N_i^c$, with $i=1,2,3$ for three generations.
Then we introduce new scalar fields, including one $SU(2)_L$ triplet $\Phi$, two $SU(2)_L$ doublets
$H$ and $H'$, and two SM singlets $S$ and $T$.
The triplet and singlet fields are introduced to generate non-standard neutrino/lepton interactions, as will be discussed later.
These particles and their quantum numbers under the $SU(3)_C \times SU(2)_L \times U(1)_Y \times U(1)_{B-L}$ gauge symmetry
are summarized in Table~\ref{Particle-Spectrum}.

\begin{table}[h]
\begin{center}

\begin{tabular}{|c|cccc|}
\hline\hline
 & $SU(3)_C$ & $SU(2)_L $ & $ U(1)_Y $&$ U(1)_{B-L}$\\
\hline
~$Q_i$~ & $\mathbf{3}$&$ \mathbf{2}$&$ \mathbf{1/6}$&$ \mathbf{1/6}$ \\
$U_i$ & $\mathbf{{3}}$&$ \mathbf{1}$&$ \mathbf{2/3}$&$ \mathbf{1/6}$ \\
~$D_i$~ & $\mathbf{{3}}$&$ \mathbf{1}$&$ \mathbf{-1/3}$&$ \mathbf{1/6}$\\
\hline
~$L_i$~ & $\mathbf{1}$&$ \mathbf{2}$&$ \mathbf{-1/2}$&$ \mathbf{-1/2}$\\
$E_i$ & $\mathbf{1}$&$ \mathbf{1}$&$ \mathbf{-1}$&$\mathbf{-1/2}$ \\
~$N_i$~ & $\mathbf{1}$&$ \mathbf{1}$&$ \mathbf{0}$&$ \mathbf{-1/2}$ \\
\hline
$XE$ & $\mathbf{1}$&$ \mathbf{1}$&$ \mathbf{-1}$&$ \mathbf{-3/2}$ \\
$XE^c$ & $\mathbf{1}$&$ \mathbf{1}$&$ \mathbf{1}$&$ \mathbf{3/2}$\\
\hline
~$\Phi$~ & $\mathbf{1}$&$ \mathbf{3}$&$ \mathbf{1}$&$ \mathbf{1}$\\
~$H$~ & $\mathbf{1}$&$ \mathbf{2}$&$ \mathbf{-1/2}$&$ \mathbf{0}$\\
~$H'$~ & $\mathbf{1}$&$ \mathbf{2}$&$ \mathbf{-1/2}$&$ \mathbf{-1}$ \\
~$S$~ & $\mathbf{1}$&$ \mathbf{1}$&$ \mathbf{0}$&$ \mathbf{-1}$\\
~$T$~ & $\mathbf{1}$&$ \mathbf{1}$&$ \mathbf{0}$&$ \mathbf{-1}$ \\
\hline\hline
\end{tabular}
\caption{Particles and their quantum numbers under the
$SU(3)_C \times SU(2)_L \times U(1)_Y \times U(1)_{B-L}$ gauge group. $T$ and $H$
sequentially develop vev that break $U(1)_{B-L}$ and $SU(2)_L\times U(1)_Y$, respectively.}
\label{Particle-Spectrum}

\end{center}
\end{table}

As we know, the  $SU(3)_C\times SU(2)_L\times U(1)_{Y}\times U(1)_{B-L}$ gauge symmetry can be obtained from the $SO(10)$ gauge symmetry breaking. However, heavy $XE$, $XE^c$ fields, and the scalars $\Phi$, $H'$, $S$, $T$ are not new particles in the traditional $SO(10)$ models. Interestingly, they can be obtained via the tensor products of the ${\mathbf 10}$ fundamental representation, as well as  ${\mathbf{16}}$ and $\overline{\mathbf{16}}$ spinor representations of $SO(10)$, {\it i.e.,} the higher representations of $SO(10)$ as follows
\begin{eqnarray}
 XE &\subset &  \overline{\mathbf{16}} \otimes \overline{\mathbf{16}} \otimes  \overline{\mathbf{16}}~,~ \nonumber \\
 XE^c &\subset & \mathbf{16} \otimes \mathbf{16} \otimes \mathbf{16} ~,~ \nonumber \\
 \Phi &\subset &  \mathbf{126} \subset \mathbf{16} \otimes \mathbf{16} ~,~ \nonumber \\
 H' &\subset &  \mathbf{10} \otimes \overline{\mathbf{16}} \otimes  \overline{\mathbf{16}}~,~ \nonumber \\
 S/T &\subset & \overline{\mathbf{120}}    \subset \overline{\mathbf{16}} \otimes \overline{\mathbf{16}} ~.~\, \nonumber
\end{eqnarray}
We consider the $SU(3)_C \times SU(2)_L \times U(1)_Y \times U(1)_{B-L}$
groups as an intermediate stage of symmetry breaking sequence after the breaking of SO(10).

The $U(1)_{B-L}$ gauge symmetry spontaneously breaks after $T$
obtains a vacuum expectation value (vev), leaving out the SM
$SU(3)_C \times SU(2)_L \times U(1)_Y$ at lower energy scales.
$H$ is the SM Higgs doublet whose vev finally breaks the electroweak gauge symmetry,
and we assume that $\Phi$, $H'$ and $S$ do not acquire vevs.
In particular, the effective Yukawa couplings
between $H'$ and charged leptons can be generated if we introduce a pair of
vector-like particles $(XE, XE^c)$ as heavy mediators and $\Phi$ can couple to lepton doublets
as well.
The $XE$ mass are generated by some UV symmetry breaking above the $U(1)_{B-L}$ scale.
 Because the CP-even neutral components of $\Phi$, $H'$, and $S$
can mix with each other, their lightest mass eigenstate $s$ can couple
to the charged leptons as well as neutrinos. In short, $T$ and $H$ are
introduced to break the $U(1)_{B-L}$ gauge symmetry and electroweak gauge
symmetry, respectively, while $\Phi$, $H'$, and $S$ are introduced to
generate the Yukawa couplings between the lightest CP-even neutral scalar and charged leptons/neutrinos.

\begin{figure}[h]
\begin{center}
\includegraphics[scale=0.5]{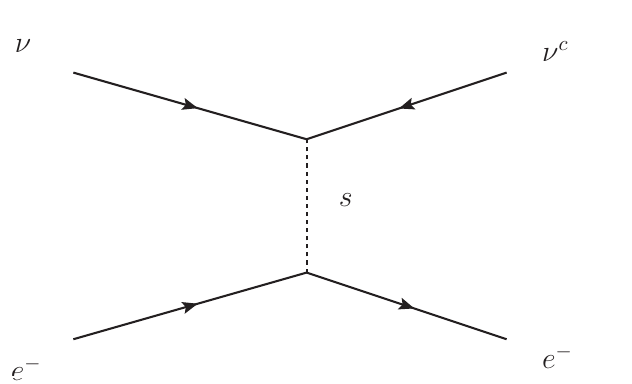}
\caption{Feynman diagram for the NSI $\nu e^- \rightarrow \nu^c e^-$ scattering.}
\label{fig:diagram}
\end{center}
\end{figure}

The Lagrangian sector involving the fermions is
\begin{eqnarray}
 -{\cal L} &=& y_{ij}^U Q_i U_j^c \overline{H} + y_{ij}^D Q_i D_j^c H + y_{ij}^E L_i E_j^c H
 + y_{ij}^{\nu} L_i N_j^c \overline{H} \nonumber\\&&
 + y_{ij}^{N} T N_i^c N_j^c
+ y_{ij}^{\Phi} L_i \Phi L_j + y_i^{H'} H' L_i XE^c \nonumber\\&&
 + y_{i}^{T} \overline{T} E_i^c XE
+ M_{XE} XE^c XE
 + {\rm H. C.}~,~\,
 \label{lag:yukawa}
\end{eqnarray}
With the $y_{ij}^{\nu} L_i N_j^c \overline{H}$ and $y_{ij}^{N} T N_i^c N_j^c$ terms,
we can generate the neutrino masses and mixings via Type I seesaw mechanism
after $T$ acquires a vev and breaks the $U(1)_{B-L}$ gauge symmetry.

Now we can see
non-standard neutrino interactions emerge from this model setup. The vector-like $(XE, XE^c)$
masses are assumed heavy as the result of UV symmetry breaking. Integrating out $(XE, XE^c)$, we can obtain
effective operators,
\begin{eqnarray}
-{\cal L} \supset -\frac{1}{M_{XE}} y_i^{H'} y_{j}^{T} H' \overline{T} L_i E_j^c + {\rm H. C.}~.~\,
\end{eqnarray}
After $U(1)_{B-L}$ gauge symmetry breaking, we get
\begin{eqnarray}
-{\cal L} \supset -\frac{\langle \overline{T} \rangle}{M_{XE}} y_i^{H'} y_{j}^{T} H' L_i E_j^c + {\rm H. C.}~.~\,
\label{Eq:ST}
\end{eqnarray}
For simplicity with phenomenology, we can assume that only $y_1^{H'}$ and $y_{1}^{T}$ are non-zero.
From the terms $y_{ij}^{\Phi} L_i \Phi L_j$ and Eq.~(\ref{Eq:ST}), we obtain
\begin{eqnarray}
  - \mathcal{L} \ \supset \ \frac{y^{\prime \Phi}_{ij}}{2} \sin\alpha \cos\beta s_1 \overline{\nu^c_{i}} \nu_{j}
+ y_e s_1 \overline{e} e
+  {\rm H. C.}~~\,
  \label{Majorana}
\end{eqnarray}
where $\nu_i$ is the neutrino mass eigenstate, $e$ is the electron, $s_1$ is a CP-even mass eigenstate from ${S, \Phi, H'}$ mixing, and
\begin{eqnarray}
y_e &=& -\frac{\langle \overline{T} \rangle}{M_{XE}} y_1^{H'} y_{1}^{T}\sin\alpha \sin\beta~.~\,
\label{eq:svv}
\end{eqnarray}
For a heavy $M_{\rm XE}$ mass much above $U(1)_{B-L}$ breaking scale the effective coupling $y_e$ can be naturally small. $\alpha, \beta$ denote the mixing angles from the neutral components in ${S, \Phi, H'}$,
\be
\left(
\begin{array}{c}
s_1 \\ s_2 \\s_3
\end{array}
\right)
=
\left(
\begin{array}{ccc}
\cos\alpha & \sin\alpha \cos\beta & \sin\alpha \sin\beta \\
-\sin\alpha & \cos\alpha \cos\beta & \cos\alpha \sin\beta \\
0 & -\sin\beta & \cos\beta
\end{array}
\right)
{\rm Re} \left(
\begin{array}{c}
S \\ \Phi^0 \\ H'^0
\end{array}
\right).
\ee
$s_1$, $s_2$, $s_3$ are their mass eigenstate and we can take $s_1$ to be the lightest state,
\begin{eqnarray}
s_1= \cos\alpha ~{\rm Re}S + \sin\alpha \cos\beta ~{\rm Re}\Phi^0 + \sin\alpha \sin\beta ~{\rm Re}H^{\prime 0}~,~\,\nonumber
\end{eqnarray}
from diagonalizing their mass matrix.
Here, we can see the triplet $\Phi$ in this model is needed to generate the effective couplings between the light mediator $s_1$ and the neutrinos, the doublet field $H'$ is needed to generate the effective couplings between $s_1$ and the charged leptons, and the singlet $S$ is needed to allow for a light mediator $s_1$.

Due to the number of scalar fields, this model has an extended scalar potential,
and its general form can be written as
\begin{eqnarray}
 V &=& m_S^2 |S|^2 -m_T^2 |T|^2-m_H^2 |H|^2 +m_{H'}^2 |H'|^2 + m_{\Phi}^2 |\Phi|^2 \nonumber\\&&
+\lambda_S |S|^4 + \lambda_T |T|^4 + \lambda_H |H|^4 + \lambda_{H'} |H'|^4
+ \lambda_{\Phi} |\Phi|^4 \nonumber\\&&
+ \lambda_{ST} |S|^2 |T|^2 + \lambda_{SH} |S|^2 |H|^2 + \lambda_{SH'} |S|^2 |H'|^2\nonumber\\&&
+\lambda_{S\Phi} |S|^2 |\Phi|^2 + \lambda_{TH} |T|^2 |H|^2 + \lambda_{TH'} |T|^2 |H'|^2 \nonumber\\&&
+\lambda_{T\Phi} |T|^2 |\Phi|^2
+ \lambda_{HH'} |H|^2 |H'|^2+\lambda_{H\Phi} |H|^2 |\Phi|^2 \nonumber\\&&
+\lambda_{H'\Phi} |H'|^2 |\Phi|^2 + \left(A_1 \Phi H H' + A_2 S \overline{H'} H
\right. \nonumber\\&& \left.
+ \lambda \Phi S H H
+{\rm H.C.}\right) ~,~
\end{eqnarray}
where $\overline{H} =i\sigma_2 H^*$ and $\overline{H'} =i\sigma_2 H^{\prime *}$ with
$\sigma_2$ the second Pauli matrix, and we neglect the $T \overline{H'} H $ and $\Phi T H H$,
which will induce the tadpole terms
for $H'$ and $\Phi$.

Here we need a light $s_1$ to
mediate non-standard soft neutrino scattering that is relevant to electron recoils.
In principle, because $S$ and $T$ carry the same quantum numbers,
without loss of generality, we can make
a $U(1)_{B-L}$ rotation so that only one linear combination of them has a vev in case they both have vevs.
After $H$ and $T$ obtain vevs, the neutral components of $\Phi$, $S$, and $H'$ fields will mix with
each other, and we assume the lightest CP-even mass eigenstate
to be very light in this paper. This typically would assume $V$ to be flat in some direction
of Re$\{S,\Phi^0,H'^0\}$.

Before analysing the electron recoil phenomenology, we would like to emphasize the difference from the
$SU(3)_C\times SU(2)_L\times U(1)_{I3R}\times U(1)_{B-L}$ model, {\it i.e.},
the traditional $U(1)_{B-L}$ model~\cite{Pati:1974yy, Marshak:1979fm, Wilczek:1979et,Mohapatra:1980qe}.
The main point is that $A_1 \Phi H H'$, $A_2 S \overline{H'} H$, and
$\lambda \Phi S H H$ terms are necessary to generate the mixings among the
neutral components of $\Phi$, $S$, and $H'$,
as well as $y_{ij}^{\Phi} L_i \Phi L_j$ to generate the
$\frac{y^{\prime \Phi}_{ij}}{2} s \overline{\nu^c_{i}} \nu_{j}$ terms.
In the traditional $SU(3)_C\times SU(2)_L\times U(1)_{I3R}\times U(1)_{B-L}$ model,
the lepton doublets $L_i$ are charged under $U(1)_{B-L}$ while the Higgs field $H$ is
neutral under $U(1)_{B-L}$. Thus, our model cannot be realized in the
traditional $SU(3)_C\times SU(2)_L\times U(1)_{I3R}\times U(1)_{B-L}$ model.

\medskip
\section{Electron recoil}
\label{sect:kinematics}

The $s_1\overline{\nu^c}\nu$ and $s_1\overline{e}e$ vertices in Eqs.\ref{Eq:ST} and~\ref{Majorana} allows $s_1$ to mediate `non-standard' electron scattering process $\nu_i e^- \rightarrow \nu^c_j e^-$ as shown in Fig.~\ref{fig:diagram}. Since $s_1$ carries lepton number, this process violates the lepton number by two units, and has no corresponding diagrams in the SM. The scattering amplitude-square is
\be
|{\cal M}|^2 = -\frac{y_\nu'^2 y_e^2(4M_e^2-t)t}{(M_s^2 -t)^2},
\ee
where $t=(p_{{\nu}^c}-p_{\nu})^\mu(p_{\nu^c}-p_{\nu})_\mu$ is the Mandelstam $t$ variable. Here we will denote $y'_\nu \equiv \frac{y^{\prime \Phi}_{ij}}{2} \sin\alpha \cos\beta$ and neglect the flavor indices for convenience. For negligible neutrino masses, $t=-2M_e E_k$ for neutrinos scattering off a free electron, and the differential cross-section is
\be
\frac{{\rm d}\sigma^{\nu e}}{{\rm d}E_k}=\frac{y'^2 y_e^2 E_k M_e(E_k+2 M_e)}{8\pi E_\nu^2(M_s^2+2 M_e E_k)^2}.
\label{eq:dsigmadEk}
\ee
$E_k$ is the electron's acquired kinetic energy after scattering, and $E_\nu$ is the incident neutrino energy, see~\ref{appendix:Cross-section} for detail. Noted the $s_1\overline{\nu^c}\nu$ vertex leads to different scattering kinematics comparing to that from a lepton number conserving $s\overline{\nu}\nu$ vertex~\cite{Cerdeno:2016sfi} at large momentum exchange, where the NSI scattering spectrum with the $s_1\overline{\nu^c}\nu$ vertex is harder at very large recoil energy $E_k\sim E_\nu$ and if $M_s$ were assumed heavy. However, since we are interested in keV recoils in this paper, the soft recoil spectra from these two types of vertices would converge and demonstrate a similar $E_k^{-1}$ behavior as long as $M_s$ is assumed smaller than the transfered momentum. In particular, it features a kinematic region:
\be
M_s\ll \sqrt{2M_e E_k}\ll M_e.
\label{eq:kinematically_soft}
\ee
Here low momentum transfer dominates the scattering as the cross-section in Eq.~\ref{eq:dsigmadEk} behaves as $d\sigma/dE_k\propto E_k^{-1}$. For $s$ mass below the keV scale, Eq.~\ref{eq:kinematically_soft} is typically satisfied by near-threshold (keV) energy transfer in electron recoil events with solar/reactor neutrinos at direct-detection experiments. Eq.~\ref{eq:dsigmadEk} is for a free electron, and the differential rate for recoil energy $E_R$ would be
\be
\frac{dN}{dE_R} = N\cdot T \cdot \epsilon(E_R) \int {\rm d}E' {\cal G}(E',E_R) \int{{\rm d}E_\nu} {\cal F}(E') \frac{{\rm d}\phi_\nu}{{\rm d} E'}\frac{{\rm d} \sigma^{\nu e}}{{\rm d} E'},
\ee
where $N$ and $T$ are the number of targets and exposure time. For comparison with XENON1T results, $\epsilon$ is the detector efficiency~\cite{Aprile:2020tmw}. ${\cal G}$ is a Gaussian smearing on $E_R$ that accounts for detector resolution,
\be
{\cal G} =\frac{1}{\sqrt{\pi}\delta_E}\exp^{-\frac{(E_R-E')^2}{\delta_E^2}}
\ee
$\delta_E=\sqrt{0.31E}+0.0037E$, where $E$ is in keV, as given in Ref.~\cite{Aprile:2020tmw}. $\phi_\nu$ is the Solar neutrino flux model that we adopt from Ref.~\cite{Bahcall:2004mz}. ${\cal F}(E)=\sum_{i}\theta(E -B_i)$ is the so-called `free electron approximation' (FEA) that serves as the electron form factor with step-functions at binding energies in the $^{54}$Xe atom as recoil thresholds. It physically represents the number of electrons that can be ionized at a given energy. FEA is adopted by XENON~\cite{Aprile:2020tmw} and it gives a 10\% correction on the $E_R\sim$keV recoil rate in our case. FEA is a popular approximation for recoil energies much higher than that of the atomic binding, and it has been shown to have very good agreement with amplitude-level photon-mediated form factor calculation for recoils at keV and above ~\cite{Chen:2016eab,Hsieh:2019hug}. We use the FEA as a reasonable approximation to the level of a few percent for $E_R> $ keV, and an amplitude-level proof is of interest for future research.

\begin{figure}[h]
\begin{center}
\includegraphics[scale=0.5]{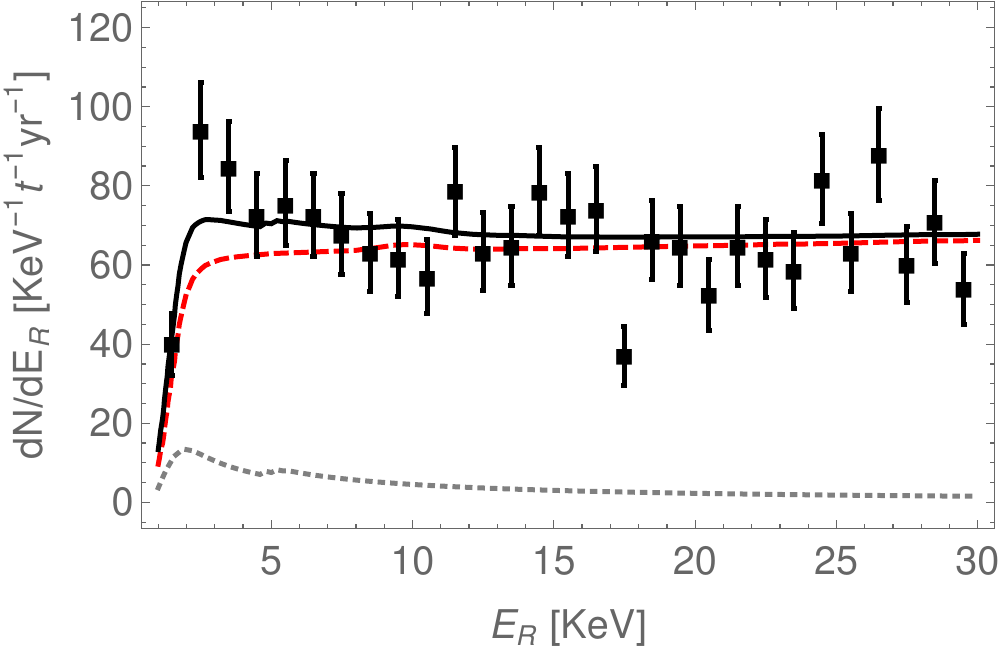}
\caption{Best-fit solar neutrino NSI event distribution (gray dotted) with $\sqrt{y_\nu y_e}={0.96}\times 10^{-6}$, background $\eta$B0 with $\eta=95.5\%$ (red dashed) and the total spectrum (solid). The NSI signal assumes the low $M_s$ limit ($M_s < $keV). }
\label{fig:fit}
\end{center}
\end{figure}

The $E_R^{-1}$ dependence from from light scalar mediated NSI leads to a relative moderate spectrum rise at the lowest energy bins, if compared to a more steep $E_R^{-2}$ dependence observed in light vector-boson mediated scenarios, as studied in Ref.~\cite{Amaral:2020tga,Boehm:2020ltd,Bally:2020yid,AristizabalSierra:2020edu}, etc. Note that photon-mediated BSM neutrino dipole interaction would also give an $E_R^{-1}$ dependence thus should yield similar explanation to data. In following sections we show the $s_1\overline{\nu^c}\nu$ NSI gives good fit the low energy electron recoils, and its significance is subject to effects from several SM radiative backgrounds.

\section{Fit to XENON1T}
\label{sect:XENON1T}

The solar neutrino's $\nu e\rightarrow \nu^c e$ event rate rises towards low energy, which is consistent with the electron recoil data reported by XENON1T. We make a likelihood fit to the 29 binned data~\cite{Aprile:2020tmw} below 30 keV, by combining these NSI-induced events with XENON1T's best-fit background modeling $B0$,
\be
\chi^2 = \sum_i \frac{(\eta B0_i + N_i^{e\nu} - N_i^{\rm data})^2}{(\delta N_i)^2}+ \frac{(1-\eta)^2}{(\delta\eta)^2},
\ee
where last term accounts for a small but crucial normalization uncertainty in the background model. In the low $E_R$ range, the detector background $B0$ is primarily the flat $^{214}$Pb component, which is a calibrated in the entire 1-210 keV range and has a 2\% statistic uncertainty. Detector efficiency modeling would contribute  another 1\% normalization uncertainty, and we take a combined $\delta\eta = 3\%$.

\begin{figure}[h]
\begin{center}
\includegraphics[scale=0.6]{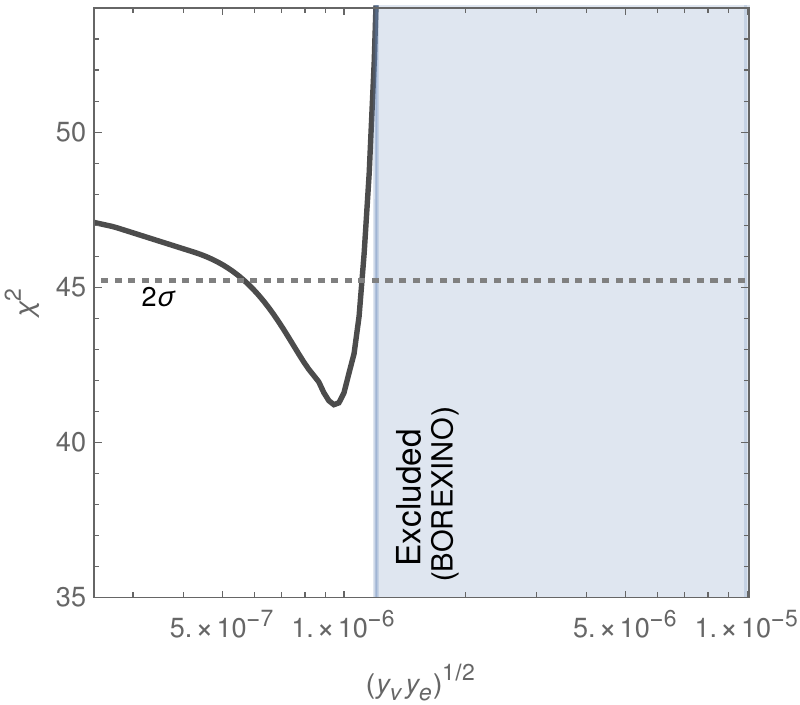}
\caption{Minimal $\chi^2$ after marginalizing over $\eta$. The $\sqrt{y_\nu y_e}\rightarrow 0$ direction approaches to the background-only fit. $5.8\times 10^{-7}<\sqrt{y_\nu y_e}<1.1\times 10^{-6}$ is a $2\sigma$ favored range around the best fit point. The shaded exclusion region is inferred from the BOREXINO bound~\cite{Borexino:2017fbd}.}
\label{fig:chi2}
\end{center}
\end{figure}

Best-fit spectra to XENON1T data is shown in Fig.~\ref{fig:fit}. Taking the low $M_s$ limit ($M_s <$ keV), a minimal $\chi^2=41$ is obtained at $\sqrt{y_\nu y_e}={0.96}\times 10^{-6}$ with the background being slightly down-scaled at $\eta-1=-4.5\%$. The best fit point yields a $\Delta \chi^2 = -6.7$ improvement over fixed $B0$ fit ($\eta=1$). The $\chi^2$ dependence on $\sqrt{y_\nu y_e}$ is plotted in Fig.~\ref{fig:chi2} and the $2\sigma$-preference threshold around the best-fit $\sqrt{y_\nu y_e}$ is shown as the dotted curve. With 28 degrees of freedom, the minimal reduced $\chi^2/{dof\#}=1.46$ corresponds to 93\% credence level (C.L.), which is not quite a perfect fit to the data below 30 keV. This is due to fluctuations above 10 keV that are still unaccounted for by the flat $^{214}$Pb background and the NSI contribution close to the detector threshold.

Notably a trace abundance of tritium below calibrated level is not ruled out as a possibility to account for the keV range recoils~\cite{Aprile:2020tmw}. The tritium induced recoil spectrum shape is similar to that from NSI and leads to degeneracy in explaining the low-energy rise. In the next session, we will show the NSI significance decreases after incorporating a more sophisticated radiative background model.

\section{Fit to PANDAX}
\label{sect:PANDAX}

After XENON1T electron recoil data was reported, the PANDAX collaboration reported their data with 100.7 ton-day exposure and 2121 events selected~\cite{Zhou:2020bvf}. There is also a rise at 3-7 keV. In contrast to the XENON experiment, more background were taken into account. Including tritium, ${ }^{127} \mathrm{Xe}$, ${ }^{85} \mathrm{Kr}$ and ${ }^{222} \mathrm{Rn}$. We also make a likelihood fit to the 24 binned data below 25 keV, by combining these NSI-induced events with PANDAX's best-fit background modeling,
\begin{eqnarray}
\chi^2 &=& \sum_{ij} \frac{(\eta_j B0_{ji}+ N_i^{e\nu} - N_i^{\rm data})^2}{(\delta N_i)^2}
+ \sum_{j}\frac{(1-\eta_j)^2}{(\delta\eta_j)^2}
\end{eqnarray}
where i denotes data point and $j=1, 2, 3, 4$ denote ${ }^{127} \mathrm{Xe}$, tritium, ${ }^{85} \mathrm{Kr}$ and ${ }^{222} \mathrm{Rn}$ background. Their best-fit value are respectively 80.8, 202.9, 1095, 735.6~\cite{Zhou:2020bvf}. $\eta$ is background floating parameters. Among this background, ${ }^{127} \mathrm{Xe}$ and tritium were considered a major factor in low energy range electron recoil rise and their statistic uncertainty are $\delta\eta_1=21\%$, $\delta\eta_2=35\%$~\cite{Wang:2020coa}. The rest of them are considered mostly flat background which are calibrated at full energy region. Their statistic uncertainty is relatively much smaller than ${ }^{127} \mathrm{Xe}$ and tritium so we don't need to consider their float and set $\eta_3=\eta_4=1$. The PANDAX detector efficiency curve is taken from~\cite{Zhou:2020}, which is degenerate with flat ${ }^{222} \mathrm{Rn}$ background so the efficiency error is not also considered. Detector smearing curve is from~\cite{Fu:2017lfc}, and we parameterized this curve as $\delta_E=0.2626\sqrt{E}+0.0426E$ where $E$ is in keV.

\begin{figure}[h]
\begin{center}
\includegraphics[scale=0.55]{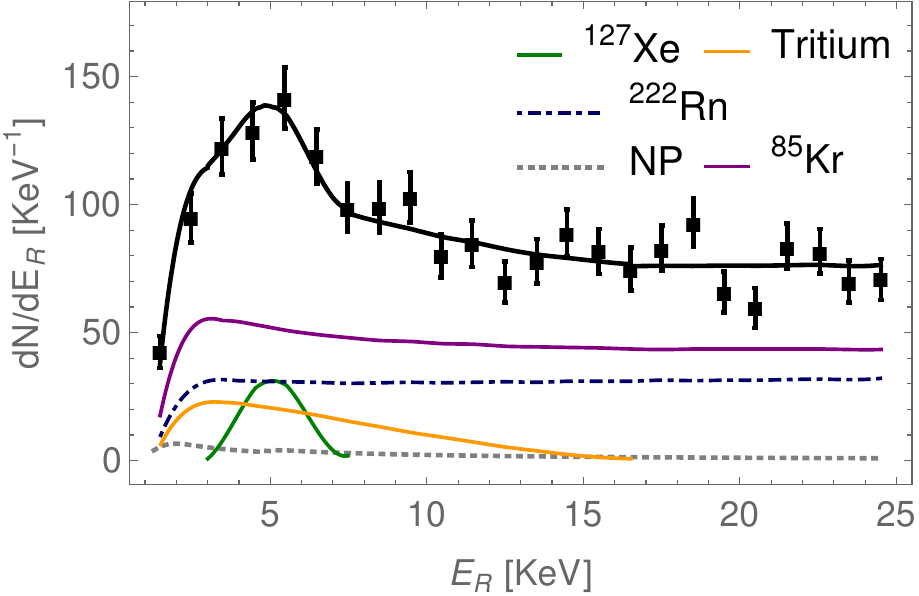}
\caption{Best-fit new physics event distribution (gray dotted) with $\sqrt{y_\nu y_e}={1.1}\times 10^{-6}$, background ${ }^{127} \mathrm{Xe}$ with $\eta_1=102.3\%$ and tritium with $\eta_2=93.3\%$. Electron recoil events are the total 100.7 ton-day data~\cite{Zhou:2020bvf}.}
\label{fig:fit_pandax}
\end{center}
\end{figure}

\begin{figure}[h]
\begin{center}
\includegraphics[scale=0.7]{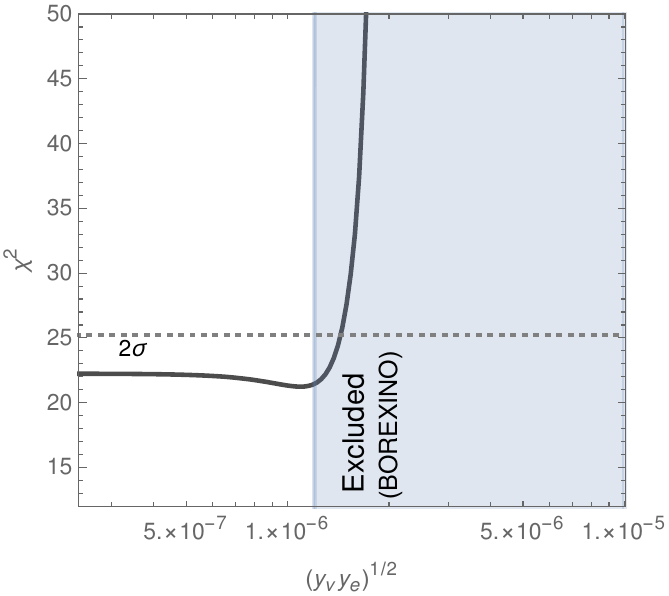}
\caption{Minimal $\chi^2$ after marginalizing over $\eta_j$. $\sqrt{y_\nu y_e}<1.4\times 10^{-6}$ is $2\sigma$ favored range and consistent with background-only fit to PANDAX data. The shaded exclusion region is inferred from the BOREXINO bound~\cite{Borexino:2017fbd}. }
\label{fig:chi2_pandax}
\end{center}
\end{figure}

Fitting PANDAX data is shown in Fig.~\ref{fig:fit_pandax}. The gray dotted line represent new physical signal with best-fit result. The background ${ }^{127} \mathrm{Xe}$ is up-scaled at $\eta_1-1=2.3\%$ and tritium is down-scaled at $\eta_2-1=-6.7\%$. The $\chi^2$ dependence on $\sqrt{y_\nu y_e}$ is plotted in Fig.~\ref{fig:chi2_pandax}. A minimal $\chi^2=21.2$ is obtained at $\sqrt{y_\nu y_e}={1.1}\times 10^{-6}$ and this point yields a $\Delta \chi^2 =-1.6$ improvement over background-only fitting results ($\chi^2=22.8$). The $2\sigma$-preference threshold around the best-fit $\sqrt{y_\nu y_e}$ is shown as the dotted curve. As PANDAX considered more background channels, the low energy range electron recoil rise can be explained well with background-only fit. Our new physical contribution is consistent with background-only fit, and we can constrain the NSI coupling to $\sqrt{y_\nu y_e}<1.4\times 10^{-6}$.

\section{Fit to XENONnT}
\label{sect:XENONnT}

Recently, XENONnT released the newest data with the larger exposure of 1.16 ton-years and lower background rate~\cite{XENON:2022mpc}. In the XENONnT experiment, they have improved systematics and achieved more than 50\% background reduction. The excess reported in XENON1T may be from trace amount of tritium and they can't confirm or exclude it at that time.  At the XENONnT experiment, they have successfully excluded the tritium component in the background model. Besides, the dominative low-energy background ${ }^{214} \mathrm{Pb}$ in XENON1T, a $\beta$-emitter from ${ }^{222} \mathrm{Rn}$, have been further reduced in XENONnT by a new high-flow radon removal system~\cite{Murra:2022mlr}. The XENONnT experiment includes nine components in the benchmark background model B0 throughout the 1-140 keV region. The main background sources below 30 keV are ${ }^{214} \mathrm{Pb}$, ${ }^{136} \mathrm{Xe}$, ${ }^{85} \mathrm{Kr}$ and materials~\cite{XENON:2022mpc}.
Unlike XENON1T, there is no obvious excess to be observed above the background at electron recoil energy below 7 keV.

Considering the major background's spectral shape is not degenerate with that of our new physical signal, here we do not need to decompose the background and will use the overall background `B0' in our analysis. We will allow this background to float by an $\eta$ parameter with an uncertainty $\delta\eta=3\%$ that corresponds to a background-only fit. Our fitting likelihood is
\be
\chi^2 = \sum_i \frac{(\eta B0_i + N_i^{e\nu} - N_i^{\rm data})^2}{(\delta N_i)^2}+ \frac{(1-\eta)^2}{(\delta\eta)^2},
\ee
which fit to XENONnT's 29 binned data below 30 keV.


\begin{figure}[h]
\begin{center}
\includegraphics[scale=0.7]{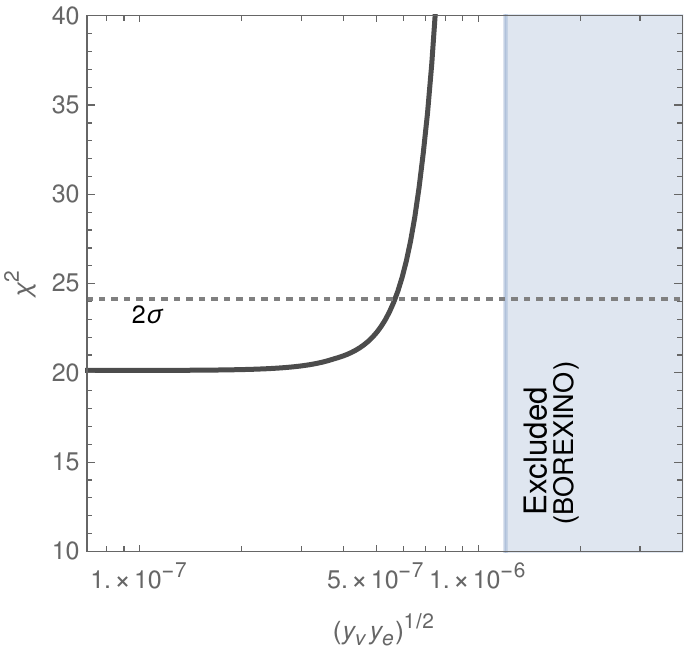}
\caption{Minimal $\chi^2$ after marginalizing over floating parameters $\eta$. $\sqrt{y_\nu y_e}<0.56\times 10^{-6}$ is $2\sigma$ favored range and consistent with background-only fit to XENONnT data. The shaded exclusion region is inferred from the BOREXINO bound~\cite{Borexino:2017fbd}. }
\label{fig:chi2_nT}
\end{center}
\end{figure}

By combining the new physical contribution and XENONnT's newest data below 30 keV from ref.~\cite{XENON:2022mpc}, we can get the corresponding limits of scalar coupling. The $\chi^2$ dependence on $\sqrt{y_\nu y_e}$ is plotted in Fig.~\ref{fig:chi2_nT}. There is no obvious minimal $\chi^2$ which is nearly a constant $\chi^2$=20.1 in the $\sqrt{y_\nu y_e}\rightarrow 0$ direction. In other words, the new physical contribution is also consistent with background-only fit. The $2\sigma$-preference threshold around the $\sqrt{y_\nu y_e}$ is shown as the dotted curve and we can get the limit of NSI coupling $\sqrt{y_\nu y_e}<0.56\times 10^{-6}$. It is clear that comparing with previous electron recoil results, XENONnT experiment gives the strongest constraint on the combined $e,\nu$ couplings to the light scalar in our model.

\section{Comparisons \& Discussion}

A few comments are due for comparing the result from lepton-number violating scattering to those existing constraints at solar and reactor neutrino experiments. As long as Eq.~\ref{eq:kinematically_soft} is satisfied, $d\sigma/dE_R\propto E_R^{-1}$ is a common kinematic feature that is also observed in lepton-number conserving ($s\overline{\nu}\nu$) scalar and neutrino magnetic dipole moment~\cite{Beacom:1999wx,Bell:2006wi,Bell:2005kz} scenarios. As most signal events are expected to be near-threshold, $s\overline{\nu}\nu$, $s\overline{\nu^c}\nu$ and $\overline{\nu}\sigma_{\mu\nu}\nu F^{\mu\nu}$ operators will lead to almost identical event distributions. This allows the scalar coupling bounds to be directly scaled from the existing $\nu_\mu$ limits. For instance, the signal events with $\nu_\mu=2.8\times 10^{-11}\mu_{\rm B}$, which corresponds to BOREXINO~\cite{Borexino:2017fbd} bounds,
shown as the shaded exclusion in Fig.~\ref{fig:chi2}. If we do not include a tritium component, the fit to XENON1T gives a slightly  ($2\sigma$) favored range $5.8\times 10^{-7}<\sqrt{y_\nu y_e}<1.1\times 10^{-6}$ just below the BOREXINO exclusion upper bound.

Given the null results from following measurements, fits to XENON1T, PANDAX, and XENONnT can also be interpreted as stringent constraints on $\sqrt{y_\nu y_e}$ that are comparable to similar experimental results. As shown in Fig.~\ref{fig:chi2}, ~\ref{fig:chi2_pandax} and ~\ref{fig:chi2_nT}, assuming null-signal the new scalar coupling $\sqrt{y_\nu y_e}$ is constrained to be below $1.1\times 10^{-6}$, $1.4\times 10^{-6}$, $0.5\times 10^{-6}$ respectively. Beside BOREXINO, there are non-standard lectron-neutrino interaction limits from several other neutrino experiments, such as GEMMA~\cite{Beda:2013mta}, TEXONO~\cite{TEXONO:2010tnr} and CHARM-II~\cite{Boehm:2004uq}, with GEMMA and BOREXINO providing the most stringent constraints. For comparison, NSI scattering with a similar $E_R^{-1}$ low-energy dependence can be interpreted from Ref.~\cite{Boehm:2020ltd,Khan:2020csx}, yielding a light scalar coupling limit $\sqrt{y_\nu y_e}\lesssim 1.2\times 10^{-6}$. We can see these limits largely fall in a very close range, with XENONnT giving the strongest constraint among these experiments, followed by the constraints is from XENON1T, GEMMA and BOREXINO, and then PANDAX.

As reactor neutrino is predominantly $\overline{\nu}_e$ at short distances, reactor neutrino constraints can be circumvented by assuming $y^\nu_{ij}$ in Eq.~\ref{Majorana} only involve $\nu_\mu$ and $\nu_\tau$, at the cost of raising the average $\sqrt{y_{\nu_\mu}y_e}, \sqrt{y_{\nu_\tau}y_e}$ by approximately 10\%-19\% as $\nu_e$ makes $1/3$ to $1/2$ of the solar neutrino flux~\cite{Agostini:2018uly} after flavor oscillation in the Sun. Note solar neutrino experimental bounds would also relax accordingly.

Since the light scalar couples to both neutrinos and electrons by effective operators at low energy scale, it can be emitted by hot electrons and neutrinos inside dense environments, hence such couplings are subject to stellar cooling constraints. For $M_s\ll$ keV, $y_\nu \le 10^{-6}$ is still consistent with stellar cooling bounds~\cite{Farzan:2002wx}. However, the electron coupling $y_e$ would be severely constrained. For instance, Ref.~\cite{Hardy:2016kme} suggested a $y_e\lesssim 10^{-14}$ bound for a scalar-electron type coupling, and similar studies~\cite{DeRocco:2020xdt} gives a $y_e\lesssim 10^{-12}$ constraint. These astrophysical bounds prevent the product $\sqrt{y_\nu y_e}$ from reaching the $10^{-6}$ level.
There are proposals to circumvent the stellar cooling bound, e.g. chameleon scenario(s) where the mediator acquires environment-dependent masses~\cite{Khoury:2003aq,Nelson:2008tn,An:2013yfc}. Realizing such scenarios requires more sophisticated model structure and is of interest for future research.

\section{Conclusion}
\label{sect:conclusion}

In this paper, we propose a model in the
$SU(3)_C\times SU(2)_L\times U(1)_{Y}\times U(1)_{B-L}$ framework that provides a light scalar
$s_1$ after $U(1)_{\rm B-L}$ and electroweak symmetry breaking, and also generates $s_1\overline{\nu^c}\nu$, $s\overline{e}e$
couplings via heavy fields above the $U(1)_{\rm B-L}$ breaking scale. These couplings violate lepton number
and lead to non-standard $\nu e\rightarrow \nu^c e$ scattering. Solar MeV neutrinos may scatter
off detector's electron via the non-standard electron-neutrino interaction, and will enhance low $E_R$ electron recoil event rate.

We calculate the recoil spectrum for Solar neutrino's $\nu e\rightarrow \nu^c e$ process, and compare the NSI spectrum in the light mediator limit ($M_s$ below keV) to other scattering processes, which share a common $E_R^{-1}$ kinematic feature near the recoil threshold energy. XENON1T data prefer a coupling range $5.8\times 10^{-7}<\sqrt{y_\nu y_e}<1.1\times 10^{-6}$, by fitting to the binned with a $3\%$ uncertainty on the $^{214}$Pb dominated background and detector efficiency. In principle this fit can be degenerate with a trace-level of tritium background. Such a $\sqrt{y_\nu y_e}$ range is consistent with existing solar neutrino measurements, and by assuming flavored $\nu_\mu,\nu_\tau$ couplings, avoids the bounds from reactor neutrino experiments, yet it is subject to rather severe $y_e$ constraint from stellar cooling, and may requires more sophisticated scenario of the mediator to comply with astrophysical limits.

The NSI-induced event rise towards the lower energy bins also allow very stringent constraints to be placed the NSI couplings. For PANDAX, the SM backgrounds including tritium and $^{127}$Xe provide excellent fit to the shape of the electron recoil spectrum. Inclusion of the light $s_1$ mediated NSI signal only yields a slight improvement and is consistent with a background only fit. PANDAX data requires $\sqrt{y_\nu y_e}<1.4\times 10^{-6}$ that is close to reactor experiment limits. In addition, the fit to XENONnT data can constrain the NSI couplings to $\sqrt{y_\nu y_e}<0.56\times 10^{-6}$, which is the strongest among terrestrial low-energy recoil and neutrino experiments.

\medskip

{\bf Acknowledgments}\\

The authors thank K. Ni, J. Ye and X. Zhou for helpful discussions with the XENON1T background modeling, and PANDAX detector efficiency and smearing. Authors also thank B. Dutta and S. Ghosh for discussions. Y.G. is partially supported by the National R\&D Program of China, 2020YFC2201601. T.L. is supported in part by the Projects 11875062 and 11947302 supported by the
National Natural Science Foundation of China, and by
the Key Research Program of Frontier Science, CAS.\\

\begin{appendix}
\section{$s\overline{\nu^c}\nu$ Scattering Amplitude}
For the $s\overline{\nu^c}\nu$ scattering on electrons, the amplitude is
\begin{equation}
i \mathcal{M}=i \frac{y_{\nu}^{\prime}y_{e}}{(p_{4}-p_{2})^{2}-M_{s}^{2}}\bar{u}(p_{4})u(p_{2})\bar{u}(p_{3})u(p_{1}) ,
\end{equation}
$p_1$, $p_3$ are the initial and recoil electron 4-momenta. $p_2$, $p_4$ are the incident and ejected neutrino 4-momenta. The amplitude-square is
\begin{equation}
\begin{aligned}
|\mathcal{M}|^{2} &=\frac{y_{\nu}^{\prime 2} y_{e}^{2}}{4(M_{s}^{2}-t)^2}\operatorname{tr}[\not p_{4}\not p_{2}]\operatorname{tr}[(\not p_{3}+M_{e})(\not p_{1}+M_{e})] \\
&=\frac{y_{\nu}^{\prime 2} y_{e}^{2}}{4(M_{s}^{2}-t)^2}(4p_{2}\cdot p_{4})(4p_{1}\cdot p_{3}+4M_{e}^{2}).
\end{aligned}
\end{equation}
Using $p_{2}\cdot p_{4}=-\frac{1}{2}t$, $p_{1}\cdot p_{3}=M_{e}^{2}+M_{e}E_{k}=M_{e}^{2}-\frac{1}{2}t$, where $t$ is the Mandelstam variable which is defined in the main text, then we get\\
\begin{equation}
|\mathcal{M}|^{2}=-\frac{y_{\nu}^{\prime 2} y_{e}^{2}\left(4M_{e}^{2}-t\right) t}{\left(M_{s}^{2}-t\right)^{2}}.
\end{equation}
In the lab frame, the total cross-section is
\begin{equation}
\sigma=\frac{1}{4M_{e}E_{\nu}}\int\frac{d^{3}\vec{p}_{3}d^{3}\vec{p}_{4}}{(2\pi)^{6}2E_{3}2E_{4}}(2\pi)^4\delta^{4}
(p_{1}+p_{2}-p_{3}-p_{4})|\mathcal{M}|^{2}.
\end{equation}
Integrating out $p_4$, we have
\begin{equation}
\sigma=\frac{1}{4M_{e}E_{\nu}}\int\frac{dE_{3}}{8\pi |\vec p_{2}|}|\mathcal{M}|^{2},
\end{equation}
$E_{3}=M_e+E_k$, $|\vec p_2|=E_{\nu}$, where $E_{k}$ is the electron's acquired kinetic energy after scattering. The differential cross-section is
\begin{equation}
\begin{aligned}
\frac{d\sigma}{dE_{k}}&=\frac{1}{4M_{e}E_{\nu}}\cdot\frac{1}{8\pi |\vec p_{2}|}\cdot|\mathcal{M}|^{2}\\
&=\frac{y^{\prime 2} y_{e}^{2} E_{k} M_{e}\left(E_{k}+2 M_{e}\right)}{8 \pi E_{\nu}^{2}\left(M_{s}^{2}+2 M_{e} E_{k}\right)^{2}},
\end{aligned}
\end{equation}
as in Eq.~\ref{eq:dsigmadEk}.
\label{appendix:Cross-section}
\end{appendix}

\bibliographystyle{elsarticle-num}
\bibliography{refs}

\end{document}